\documentclass[preprint,10pt,5p]{elsarticle}
\pdfoutput=1 
\RequirePackage[colorlinks,breaklinks,pdftitle={ATLAS paper},pdfauthor={The ATLAS Collaboration}]{hyperref}  
\usepackage{graphicx}
\usepackage{overpic}
\usepackage{amssymb}
\DeclareGraphicsExtensions{.pdf}
\biboptions{compress}

\usepackage{atlasphysics}
\usepackage{zptdefs}

\journal{Physics Letters B}

\begin{document}
\begin{frontmatter}

\title{\vspace{-0.8cm} \flushright{\normalsize{CERN-PH-EP-2011-095}} \\ 
\center{Measurement of the transverse 
momentum distribution of \Zg\ bosons in proton-proton collisions at
$\sqrt{s} =7$~TeV with the ATLAS detector}
}

\author{The ATLAS Collaboration}

\date{October 5, 2011}

\begin{abstract}
A measurement of the \Zg\ transverse momentum (\ptz) 
distribution
in proton-proton collisions at $\rts = 7$~\TeV\ is presented using \Zgee\ and \Zgmm\ decays
collected with the ATLAS detector in data sets with integrated luminosities of
35~\ipb\ and 40~\ipb, respectively. The normalized differential 
cross sections are measured separately for electron and muon decay channels 
as well as for
their combination up to $\ptz$ of $350\GeV$ for invariant dilepton masses $66\GeV<m_{\ell\ell}<116\GeV$.
The measurement is compared to predictions of perturbative QCD and various event generators.
The prediction of resummed QCD combined with fixed order perturbative QCD is found to be in good agreement
with the data.
\end{abstract}

\begin{keyword}
Z bosons\sep
Differential cross section\sep
Perturbative QCD\sep
Event generators\sep
Monte Carlo models
\end{keyword}

\end{frontmatter}

\global\emergencystretch = .1\hsize

\section{Introduction}

In hadron collisions, the weak vector bosons $W$ and $Z$ are produced 
with a momentum component transverse to the beam axis, which is
balanced by a recoiling hadronic system mainly arising from initial 
state QCD radiation of quarks and gluons.
The measurement of the boson transverse momentum offers a very sensitive way
of studying dynamical effects of the strong interaction, complementary to
measurements of the associated production of the bosons with jets~\cite{Aad:2010pg}.
The simple signatures of \Zgee\ and \Zgmm\ production, which can be
identified with little background, enable a precise measurement of the
boson transverse momentum (\ptz) and thus provide
an ideal testing ground for predictions of QCD and phenomenological models.
Moreover, the knowledge of the  \ptz\ distribution is
crucial to improve the modelling of $W$ boson production needed for a
precise measurement of the $W$ mass~\cite{Aaltonen:2007ps,Abazov:2009cp}, in particular in the low \ptz\ region which
dominates the cross section.

This Letter presents a measurement of the \Zg\ normalized transverse momentum
distribution
in proton-proton collisions at $\rts = 7$~\TeV\ using \Zgee\ and \Zgmm\ decays
collected in 2010 with the ATLAS detector in data sets with integrated luminosities of
35~\ipb\ and 40~\ipb, respectively.
In the normalized transverse momentum distribution, many systematic
uncertainties cancel. In particular, the precision of the measurement is
not impaired by the uncertainty on the integrated luminosity.
The normalized transverse momentum distribution is defined in the following way:
$1/\sigma^{\mathrm{fid}} \times \der\sigma^{\mathrm{fid}}/\der \ptz,$
where $\sigma^{\mathrm{fid}}$ is the measured inclusive cross section of 
$pp\ra\Zg+X$ multiplied by the branching ratio of $\Zg\ra\ell^{+}\ell^{-}$ within the fiducial
acceptance, and $X$ denotes the underlying event and the recoil system.
For both the \ee\ and \mm\ channels,
the fiducial acceptance is defined by the lepton transverse
momentum and pseudorapidity\footnote{The
nominal $pp$ interaction point at the centre of the detector is
defined as the origin of a right-handed coordinate system. The
positive $x$-axis is defined by the direction from the interaction
point to the centre of the LHC ring, with the positive $y$-axis
pointing upwards. The azimuthal angle $\varphi$ is measured around the
beam axis and the polar angle $\theta$ is the angle from the
$z$-axis. The pseudorapidity is defined as $\eta = -\ln
\tan(\theta/2)$.}, and by the invariant mass of the lepton pair $m_{\ell\ell}$:
$\pt^\ell>20\GeV$, $|\eta^\ell|<2.4$ and $66\GeV<m_{\ell\ell}<116\GeV$.
The measurements in both decay channels are corrected for detector effects and QED
final state radiation (FSR). The combined result is compared to
predictions of perturbative QCD calculations and QCD-inspired models implemented in
various event generators.

Predictions and previous measurements are discussed in Section~\ref{sec:Predictions}.
The ATLAS detector and trigger are described in Section~\ref{sec:Detector}.
In Sections~\ref{sec:Simulation} and~\ref{sec:Selection} the event simulation and
selections are described.
Section~\ref{sec:Measurement} reports the \ptz\ measurements for different treatments
of QED final state radiation.
Systematic uncertainties are discussed in Section~\ref{sec:Systematics}.
Section~\ref{sec:Results} presents the combined
result which is compared with various models.

\section{QCD predictions and previous measurements}
\label{sec:Predictions}

Perturbative QCD (pQCD) calculations have been performed
up to $\mathcal{O}(\alphas^2)$ in the strong coupling constant
\alphas\ \cite{Arnold:1988dp,Gonsalves:1989ar} and are expected to be reliable at large
\ptz. In this kinematic regime, the cross section is dominated by the radiation of a
single parton.
Fully differential inclusive boson production
cross sections can be obtained at $\mathcal{O}(\alphas^2)$ with
{\sc Fewz}~\cite{Melnikov:2006kv,Gavin:2010az} and
{\sc Dynnlo}~\cite{Catani:2009sm}.
While the integrated $\mathcal{O}(\alphas^2)$ cross section predictions are
finite,
the fixed order pQCD prediction diverges 
at vanishing \ptz. In this regime, the leading contribution of
multiple soft gluon emissions to the inclusive cross section 
can be resummed to all orders~\cite{Davies:1984hs,CSS,Ladinsky:1993zn} up to
next-to-next-to-leading logarithms (NNLL)~\cite{Bozzi:2010xn} in \alphas.
The {\sc Resbos} generator~\cite{Balazs:1997xd} matches the prediction of
soft gluon resummation including a non-perturbative form 
factor~\cite{Landry:2002ix} at low \ptz\ with the fixed order pQCD
calculation at $\mathcal{O}(\alphas)$ at high \ptz, which is corrected to
$\mathcal{O}(\alphas^2)$ using $K$-factors.

Similar to resummed calculations, parton showers provide an all-order
approximation of parton radiation in the soft and collinear region.
In order to describe the large \ptz\ region, the parton shower based
leading-order event generators {\sc Pythia}~\cite{pythia} and
{\sc Herwig}~\cite{herwig} apply
weights to the first or hardest branching, respectively, to effectively
merge the $\mathcal{O}(\alphas^0)$ and $\mathcal{O}(\alphas)$ pQCD predictions.
The next-to-leading order (NLO) Monte Carlo generators
\mcatnlo~\cite{mcatnlo} and {\sc Powheg}~\cite{powheg}
incorporate NLO QCD matrix elements consistently into the parton shower
frameworks of {\sc Herwig} or {\sc Pythia}.

The {\sc Alpgen}~\cite{Mangano:2002ea} and {\sc Sherpa}~\cite{Gleisberg:2008ta}
event generators
implement tree-level matrix elements for the generation of multiple hard partons
in association with the weak boson. The matrix-element calculations for
various parton multiplicities are matched with parton showers (which in the case
of {\sc Alpgen} are provided by either {\sc Pythia} or {\sc Herwig}) such that
double counting is explicitly avoided by means of weighting
procedures~\cite{Krauss:2002up} or veto algorithms~\cite{Mangano:2002ea}.

As the predictions of these generators differ significantly and
show a considerable dependence on adjustable internal
parameters~\cite{Alwall:2007fs}, a precise measurement of the boson
transverse momentum distribution is an important input to validate and tune these
models.

The \Zg\ boson \pt\ distribution has been measured in 
proton-antiproton collisions at the Tevatron collider at centre of mass 
energies of $\rts=1.8$~\TeV\ and $1.96$~\TeV\ 
\cite{Affolder:1999jh,Abbott:1999yd,Abazov:2007nt,Abazov:2010kn,Abazov:2010mk}. 
For $\ptz\lesssim 30$~\GeV, these measurements found a good agreement 
with {\sc Resbos} and disfavoured models~\cite{Berge:2004nt} suggesting 
a broadening of the
\ptz\ distribution for small $x$ values~\cite{Abazov:2007nt,Abazov:2010mk}, where $x$ is 
the fraction of the momentum of one of the two partons with respect to the proton momentum. 
At large \ptz, the $\mathcal{O}(\alphas^2)$
pQCD prediction was reported to underestimate the measured 
cross section by up to about
25\%~\cite{Abazov:2007nt,Abazov:2010kn}.

\section{The ATLAS detector}
\label{sec:Detector}

The ATLAS detector system~\cite{DetectorPaper:2008} comprises 
an inner tracking detector immersed in a $2$~T axial magnetic field,
a calorimeter, and a large muon spectrometer with a superconducting toroid 
magnet system.
Charged particle tracks and vertices are reconstructed with silicon pixel and 
strip detectors covering $|\eta|<2.5$ and transition radiation 
detectors covering $|\eta| < 2.0$.
These tracking detectors are surrounded by a finely segmented calorimeter
system which provides three-dimensional reconstruction of particle showers
up to $|\eta| < 4.9$.
The electromagnetic compartment uses liquid argon as the active material
and is divided into barrel ($|\eta| < 1.5$),
end-cap ($1.4 < |\eta| < 3.2$) and forward ($3.1 < |\eta|< 4.9$) components.
The hadron calorimeter is based on scintillating tiles in the central
region ($|\eta| < 1.7$). It is extended up to $4.9$ in
pseudorapidity by end-caps and forward calorimeters which use liquid argon.
The muon spectrometer is based on three large superconducting
toroids arranged with an eight-fold azimuthal coil symmetry around
the calorimeters, covering a range of $|\eta|<2.7$ and providing an
integral magnetic field varying from about 1~to $8\Tm$.
Three stations of drift tubes and cathode strip chambers enable 
precise muon track measurements, and resistive-plate and thin-gap
chambers provide muon triggering capability and additional
measurements of the $\varphi$ coordinate.

The ATLAS detector has a three-level trigger system which 
reduces the event rate to approximately 200~Hz before data transfer to
mass storage. 
The triggers employed require the presence of a single
electron or muon candidate with $\pt>15$~\GeV\ or $\pt>13$~\GeV, respectively. 
Lower thresholds were used for the early data.
The trigger efficiencies are defined for \Zgee\ and \Zgmm\ events as the
fraction of triggered electrons or muons with
respect to the reconstructed lepton and are studied as a function of their
\pt\ and $\eta$. For muons, the efficiencies are also obtained separately in
$\varphi$ regions which match the geometry of the trigger chambers.
The efficiency for single leptons is derived from data using
\Zgll\ candidate events or using independent triggers by
matching reconstructed lepton candidates to trigger
signals in the calorimeter (muon spectrometer) in case of the \ee\ (\mm) decay channel.
For $\pt^\ell>20$~\GeV, the efficiency is  99\% for electrons and 77\% (93\%) for muons in the
barrel (end-cap).
For signatures with two high-\Et\ electrons, the trigger is fully efficient.
The trigger efficiency for \Zgmm\ events is determined to be 
on average 97.7\% and to be constant as a function of \ptz\ within an 
uncertainty of $0.1-0.7\%$ depending on \ptz.

\section{Event simulation} 
\label{sec:Simulation}

The properties, including signal
efficiencies and acceptances, of \Zgee, \Zgmm\ and background processes
are modelled with {\sc Pythia}~\cite{pythia} using the MRST2007LO$^*$~\cite{LOmod} parton distribution functions (PDF), 
\mcatnlo~\cite{mcatnlo} and {\sc{}Powheg} using  CTEQ6.6~\cite{Nadolsky:2008zw} PDFs.
\mcatnlo\ uses {\sc{}Herwig} for the parton shower and {\sc{}Jimmy}~\cite{Butterworth:1996zw} 
for the underlying event. {\sc{}Powheg} is interfaced to {\sc{}Pythia} for the underlying event and 
the parton shower.
The event
generators are interfaced to {\sc Photos}~\cite{photos} to simulate QED FSR.
Version 6.4 of {\sc Pythia} is used with the \pt-ordered parton shower and with
parameters describing the properties of the underlying event which were tuned to 
Tevatron measurements~\cite{ATL-PHYS-PUB-2010-002}.
For systematic studies and comparisons, a \mcatnlo\  based signal
sample is used with underlying event parameters  ({\sc{}Jimmy}) tuned to Tevatron
and $7\TeV$ ATLAS $pp$ collision data~\cite{mc10Herwig}.
The response of the ATLAS detector to the generated particles is modelled using 
{\sc Geant4}~\cite{geant4}, and the fully simulated events~\cite{aad:2010wqa} are passed through
the same reconstruction chain as the data.
The Monte Carlo simulation (MC) 
is corrected for differences with respect to the data in the lepton reconstruction and identification
efficiencies as well as in energy (momentum) scale and resolution.
The efficiencies are determined from a tag-and-probe
method based on reconstructed $Z$ and $W$ 
events~\cite{WZXsInclusive,ATLAS:eperf},
while the resolution and scale
corrections are obtained from a fit to the observed $Z$ boson line shape.
The lepton identification efficiencies can depend on the hadronic activity, which is correlated
with the \Zg\ transverse momentum. Therefore, using the tag-and-probe method,
it is verified that the 
\ptz\ dependence of the single lepton efficiency is correctly modelled after efficiency corrections. Differences between data and
simulation are mostly consistent with statistical fluctuations and are considered as systematic 
uncertainties due to the modelling of the efficiencies as described in Section~\ref{sec:Systematics}.

Multiple $pp$ interactions per bunch crossing ({\em pileup}\/) are accounted for by 
overlaying simulated minimum bias events. To match the observed 
instantaneous luminosity profile, the MC events are reweighted to
yield the same distribution of the number of primary vertices as measured in
the data.

In the following, if not stated otherwise, the generated samples are fully simulated,
pileup and resolution corrected, and interfaced to {\sc{}Photos},
with {\sc{}Pythia} as the default MC signal sample. The background samples are produced 
with {\sc{}Pythia} for $W\rightarrow\ell\nu$, with {\mcatnlo} and {\sc{}Powheg} for $t\bar{t}$, 
and with {\sc{}Pythia} and {\mcatnlo} for $\Zg\ra\tau^{+}\tau^{-}$.

\begin{figure*}[t]
  \centering
  \begin{overpic}[width=0.49\textwidth]{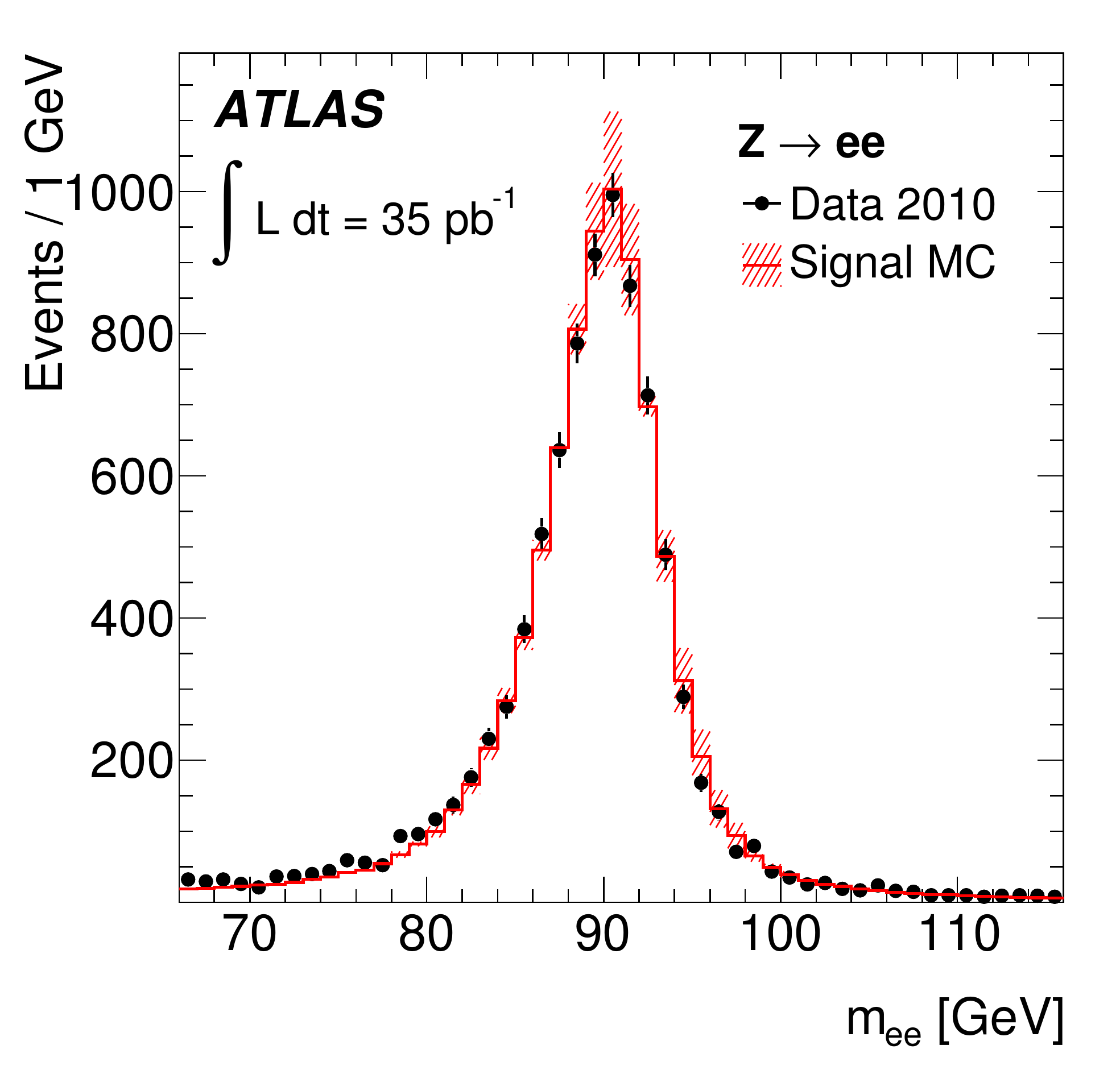}
    \put(84,84.5){\large(a)}
  \end{overpic}
  \begin{overpic}[width=0.49\textwidth]{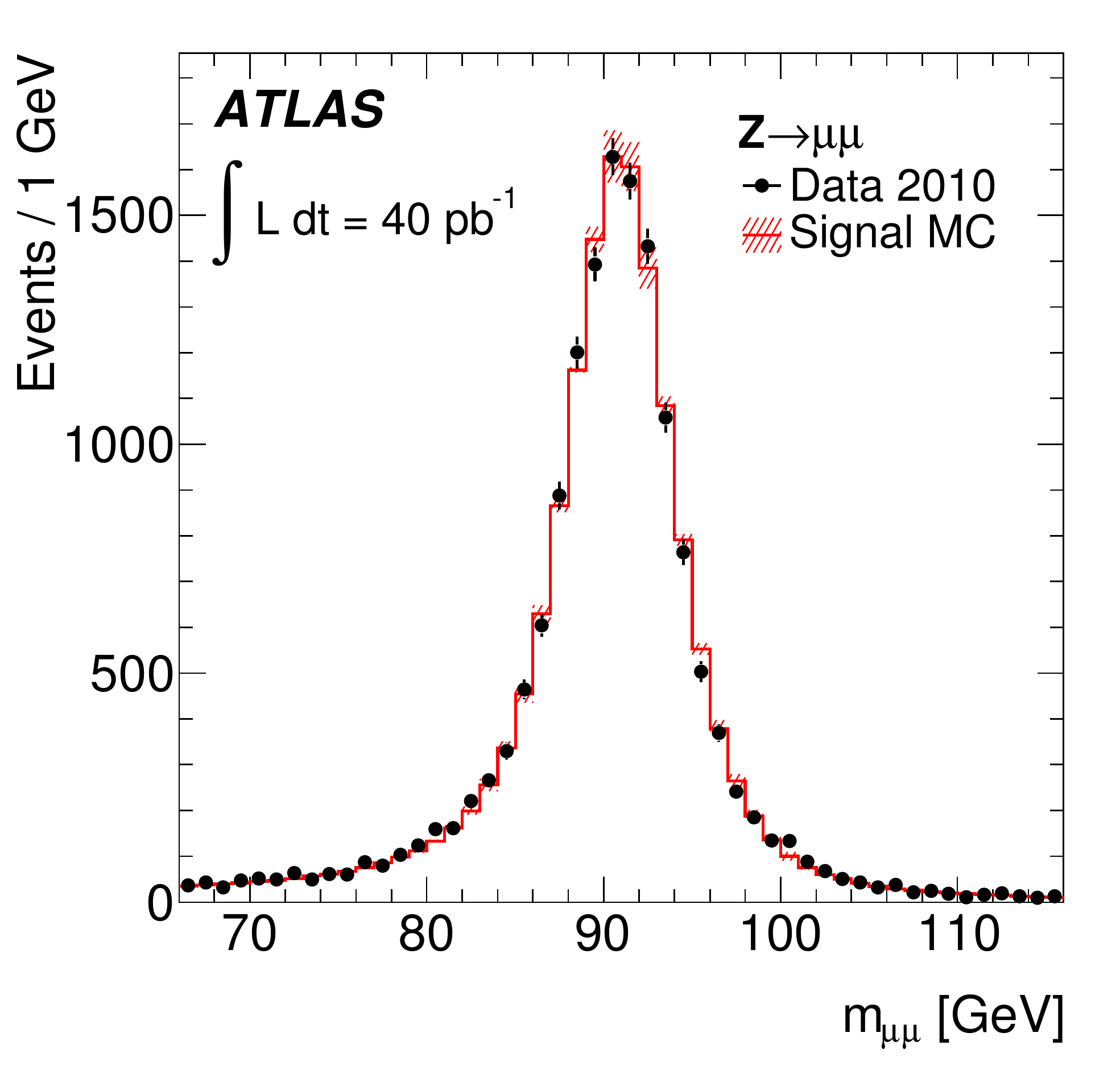}
    \put(84,84.5){\large(b)}
  \end{overpic}
  \caption{
    The observed (a) dielectron and (b) dimuon invariant mass
    distributions compared to simulation. The shaded bands indicate the
    systematic uncertainty on the MC prediction due to the lepton momentum resolution correction.
    The {\sc Pythia} sample, which is normalized to the next-to-next-to-leading order (NNLO) 
    prediction~\cite{zwprod,Melnikov:2006kv,Gavin:2010az}, is used for signal events. The total background, which is invisible on this scale,
    amounts to only $1.5\%$ and $0.4\%$ in the \ee\ and \mm\ decay channels, respectively.
  \label{fig:mll}
  }
\end{figure*}

\section{Event selection}
\label{sec:Selection}
The analysis uses data taken during stable beam conditions
with properly operating inner detector, magnets, and calorimeter or muon
spectrometer, in case of the \ee\ or \mm\ channel, respectively.
Events are required to have at least one primary vertex 
reconstructed from at least three tracks. \Zg\ events are
selected by requiring two oppositely charged electrons or muons, 
defined below, with an invariant mass $66\GeV < \mass{\ell} < 116\GeV$.

Electrons are reconstructed from the energy deposits in the calorimeter matched
to inner detector tracks.
They are required to have
a transverse energy $\Et^e > 20$~\GeV\ and pseudorapidity $|\eta^e|<2.4$, excluding the
transition regions between the barrel and end-cap calorimeter components at
$1.37 < |\eta^e| < 1.52$. They should pass {\em medium} identification criteria
based on shower shape and track quality variables~\cite{WZXsInclusive} to
provide rejection against hadrons. The single electron selection efficiency varies in
the range $90-96\%$ depending on pseudorapidity and azimuthal angle.

Muons are reconstructed from matching tracks in the inner detector and muon
spectrometer with $\pt^\mu > 20$~\GeV\ and $|\eta^\mu|<2.4$
measured using the information from these two detector subcomponents.
To ensure the compatibility of the muon track with the primary vertex, the corresponding 
impact parameters in the transverse and longitudinal direction with respect to the beam axis
have to be smaller than $1$~mm and $5$~mm, respectively.
The muon candidates are required to be isolated to suppress background 
from heavy flavour production using the transverse momentum sum of tracks 
with $\pt > 1$~\GeV\ within a cone of size
$\Delta R = \sqrt{(\Delta \eta)^2 + (\Delta \varphi)^2}=0.2$
around the muon track. This sum has to be smaller than
$0.2 \times \pt^\mu$.
The single muon selection efficiency is of the order of $87-95\%$ depending on 
pseudorapidity and azimuthal angle.

After this selection, 8\,923 \Zgee\ and 15\,060
\Zgmm\ candidate events are found in the data. 
The main reasons for the difference in the number of candidate events in the \ee\
and \mm\ decay channels are:  the integrated luminosity for
which all relevant detector components were operating properly; regions in the
electromagnetic calorimeter with
readout problems; a reduced acceptance for electrons in
the transition region between the barrel and end-cap calorimeter.
Figure~\ref{fig:mll} demonstrates the agreement of the data and the simulation in the dilepton mass
spectrum for the selected events including the background, after
applying the corrections described in Section~\ref{sec:Simulation}.

The background contribution from \Zgtt, $W$ boson, and $t\bar{t}$
production is estimated as a function of \ptz\ using simulation, where the cross sections 
are normalized to next-to-next-to-leading order (NNLO) predictions 
for \Zg\ and $W$ and NLL-NLO
predictions for $t\bar{t}$ production. The procedure outlined in 
Ref.~\cite{WZXsInclusive} is followed here.

For both \ee\ and \mm\ channels, 
the main background at high \ptz\ arises
from $t\bar{t}$ production.
At low \ptz\ the background is dominated by 
QCD multijet production, where a jet is falsely identified as a 
primary $e$ or $\mu$.
Its contribution is determined from the data as follows.

In the \ee\ channel, the normalization of the multijet contribution is
derived from a fit of signal and background templates to the observed
dilepton invariant mass distribution with loosened identification
requirements for one of the two reconstructed electron candidates. An
extended mass range of $50\GeV < m_{ee} < 130\GeV$ is used which provides a
better background constraint in the off-resonance region. The normalization derived from 
this loosened selection has to be scaled to the \Zg\ event selection, which requires two reconstructed
electron candidates of medium quality. The scaling factor is determined
from a QCD multijet enhanced control sample with single electron candidates
which fulfil the loosened electron identification requirements. The
contamination with other events, in particular the
contribution from $W$ production, is suppressed by rejecting events with
large missing transverse energy. The remaining contamination is determined
using simulated events.
The systematic uncertainty of the
normalization is determined by varying the background templates and the
criteria for the loosened selection. The shape of the multijet background as
a function of \ptz\ is determined from a dielectron sample with an
invariant mass $66\GeV<m_{ee}<116\GeV$ for which exactly one electron passes
and one fails the medium identification criteria. The difference between
same and opposite sign events is taken as the shape uncertainty.

For the \mm\ selection, the multijet contribution is estimated from
four two-dimensional regions which are obtained by changing the mass window 
to $40\GeV <
m_{\mu\mu} < 60\GeV$ or by inverting the isolation criterion. The normalization
is determined from the number of candidate events in these four regions
which is corrected by the expected number of non-QCD multijet background events and number 
of signal events.
The number of signal events is determined from the measurement in the signal
region which is corrected for the background including the multijet 
background, and 
extrapolated to the other regions using relative efficiencies extracted from the simulation.
The resulting equation can be solved for the number of QCD multijet background 
events.
The shape is determined from the control region where both muon candidates are
non-isolated and their invariant mass is within $66\GeV < m_{\mu\mu}
< 116\GeV$. Systematic uncertainties are determined by comparing the results to those of 
alternative methods which use template fits or same sign and opposite sign events.

The total background contribution, which amounts to $(1.5 \pm
0.6)\%$ and $(0.4\pm0.2)\%$ in the \ee\ and \mm\ channels, respectively, is
found to increase from 0.5\%~(0.2\%) to 3.5\%~(1.5\%) as a function of \ptz\
in the \ee~(\mm) analysis.

\section{The measured \ptz\ distribution}
\label{sec:Measurement}
The \Zg\ transverse momentum is reconstructed from the measured lepton
momenta.
The \ptz\ range is divided into 19 bins from $0\GeV$ to $350\GeV$ with
widths of $3\GeV$ for $\ptz < 30\GeV$ and increasing widths at larger
transverse momenta as given in Table~\ref{tab:xsect_emu}.
For this binning, the fraction of simulated events reconstructed in a particular \ptz\ bin which have
generator-level \ptz\ in the same bin is always better than 60\% and reaches values above
90\% in the highest \ptz\ bins for both decay channels.
The bin-by-bin efficiency, defined as the ratio 
between the number of signal events which pass the final selection and the 
total number of generated events within the fiducial region, are on 
average about 56\% in the \ee\ and 83\% in the \mm\ channel, 
respectively.
Three \Zg\ candidate events with $\ptz>350\GeV$ are found, which
are not considered further due to the limited statistical significance.

The observed \ptz\ spectrum is found to be well described by the simulation,
using the default MC signal samples
and background estimations as described in Sections~\ref{sec:Simulation} and~\ref{sec:Selection}.
A bin-by-bin efficiency correction is used to correct (unfold) the observed data for detector effects and 
QED FSR, where 
the correction factors are determined from the
default MC signal sample. Alternative matrix unfolding
methods~\cite{MR99a:65037,Kaipio:1338003},
which explicitly take the bin-to-bin migration into account,
yield compatible results. 
However, these techniques require higher data statistics to fully exploit
their advantages. Therefore they are currently used only to estimate the
systematic uncertainties 
due to the unfolding method as discussed in Section~\ref{sec:Systematics}.

\begin{table*}[tb]
\caption{
The measured normalized differential cross section
$1/\sigma^{\mathrm{fid}}\der\sigma^{\mathrm{fid}}/\der \ptz$ in bins of \ptz\
for \Zgee\ and \Zgmm\ production. The cross sections, which are to be
multiplied by the factor $k$, are reported with respect to three different
treatments of QED final state radiation for the definition of lepton and
\Zg\ momentum at particle level: corresponding to the \Zg\ propagator (propag.),
to dressed leptons (with
recombined radiated photons within a cone of $\Delta R=0.1$), and to bare
leptons, respectively.
The relative statistical (stat.) and total systematic (syst.) uncertainties are given.
The data can be obtained electronically through the HepData repository~\cite{hepdata}.
\label{tab:xsect_emu}
}
\begin{center}
  \begin{tabular}{c@{-}crrrrrrcrrrrrr}
\hline\noalign{\smallskip}
\multicolumn{2}{c}{\ptz\ bin} & 
\multicolumn{12}{l}{$1/\sigma^{\mathrm{fid}} \der\sigma^{\mathrm{fid}}/\der \ptz$ (\GeV$^{-1}$)} \\[\smallskipamount]
\multicolumn{2}{c}{(\GeV)} & \multicolumn{3}{l}{\Zgee} & & \multicolumn{2}{l}{uncert. (\%)}  & &
\multicolumn{3}{l}{\Zgmm}  & &\multicolumn{2}{l}{uncert. (\%)} \\
\cline{3-8}\cline{10-15}
\multicolumn{2}{c}{}  & 
\multicolumn{1}{l}{propag.} & \multicolumn{1}{l}{dressed} & \multicolumn{1}{l}{bare} 
 & \multicolumn{1}{l}{$k$}
 & \multicolumn{1}{l}{stat.} & \multicolumn{1}{l}{syst.} & &
\multicolumn{1}{l}{propag.} & \multicolumn{1}{l}{dressed} & \multicolumn{1}{l}{bare} 
 & \multicolumn{1}{l}{$k$} 
& \multicolumn{1}{l}{stat.} & \multicolumn{1}{l}{syst.} \\
\hline\noalign{\smallskip}
0 & 3 & 3.48 & 3.40 & 3.21 & $\cdot 10^{-2}$ & 3.3 & 4.7 &   & 3.75 & 3.66 & 3.58 & $\cdot 10^{-2}$ & 2.6 & 5.0\\
3 & 6 & 5.85 & 5.78 & 5.60 & $\cdot 10^{-2}$ & 2.4 & 3.3 &   & 5.81 & 5.74 & 5.68 & $\cdot 10^{-2}$ & 2.0 & 4.0\\
6 & 9 & 4.61 & 4.62 & 4.64 & $\cdot 10^{-2}$ & 2.7 & 2.3 &   & 4.67 & 4.68 & 4.69 & $\cdot 10^{-2}$ & 2.1 & 1.6\\
9 & 12 & 3.43 & 3.46 & 3.56 & $\cdot 10^{-2}$ & 3.1 & 2.4 &   & 3.50 & 3.54 & 3.58 & $\cdot 10^{-2}$ & 2.4 & 1.6\\
12 & 15 & 2.93 & 2.97 & 3.09 & $\cdot 10^{-2}$ & 3.3 & 2.7 &   & 2.67 & 2.72 & 2.76 & $\cdot 10^{-2}$ & 2.8 & 1.7\\
15 & 18 & 2.04 & 2.08 & 2.16 & $\cdot 10^{-2}$ & 3.9 & 3.0 &   & 2.13 & 2.17 & 2.20 & $\cdot 10^{-2}$ & 3.1 & 1.7\\
18 & 21 & 1.64 & 1.67 & 1.73 & $\cdot 10^{-2}$ & 4.4 & 3.3 &   & 1.69 & 1.72 & 1.74 & $\cdot 10^{-2}$ & 3.5 & 1.8\\
21 & 24 & 1.32 & 1.33 & 1.37 & $\cdot 10^{-2}$ & 4.8 & 3.6 &   & 1.35 & 1.36 & 1.37 & $\cdot 10^{-2}$ & 4.0 & 1.8\\
24 & 27 & 1.08 & 1.08 & 1.11 & $\cdot 10^{-2}$ & 5.5 & 3.8 &   & 1.15 & 1.16 & 1.17 & $\cdot 10^{-2}$ & 4.3 & 1.9\\
27 & 30 & 1.02 & 1.03 & 1.03 & $\cdot 10^{-2}$ & 6.5 & 4.0 &   & 0.87 & 0.88 & 0.88 & $\cdot 10^{-2}$ & 5.0 & 2.0\\
30 & 36 & 7.22 & 7.24 & 7.26 & $\cdot 10^{-3}$ & 4.8 & 4.2 &   & 6.45 & 6.46 & 6.45 & $\cdot 10^{-3}$ & 4.1 & 2.1\\
36 & 42 & 4.89 & 4.88 & 4.85 & $\cdot 10^{-3}$ & 5.8 & 4.5 &   & 4.63 & 4.63 & 4.62 & $\cdot 10^{-3}$ & 4.9 & 2.2\\
42 & 48 & 3.66 & 3.64 & 3.59 & $\cdot 10^{-3}$ & 7.0 & 4.8 &   & 3.97 & 3.95 & 3.94 & $\cdot 10^{-3}$ & 5.3 & 2.4\\
48 & 54 & 3.26 & 3.25 & 3.20 & $\cdot 10^{-3}$ & 7.8 & 5.0 &   & 2.90 & 2.88 & 2.86 & $\cdot 10^{-3}$ & 6.2 & 2.6\\
54 & 60 & 2.14 & 2.13 & 2.08 & $\cdot 10^{-3}$ & 9.2 & 5.4 &   & 2.14 & 2.13 & 2.11 & $\cdot 10^{-3}$ & 7.2 & 2.7\\
60 & 80 & 1.21 & 1.20 & 1.17 & $\cdot 10^{-3}$ & 6.5 & 5.7 &   & 1.31 & 1.30 & 1.28 & $\cdot 10^{-3}$ & 5.1 & 3.0\\
80 & 100 & 5.69 & 5.63 & 5.44 & $\cdot 10^{-4}$ & 9.8 & 5.9 &   & 5.52 & 5.47 & 5.40 & $\cdot 10^{-4}$ & 7.8 & 3.5\\
100 & 180 & 1.74 & 1.73 & 1.67 & $\cdot 10^{-4}$ & 9.6 & 6.1 &   & 1.52 & 1.51 & 1.49 & $\cdot 10^{-4}$ & 7.5 & 4.4\\
180 & 350 & 0.78 & 0.77 & 0.73 & $\cdot 10^{-5}$ & 27.0 & 7.8 &   & 1.14 & 1.14 & 1.11 & $\cdot 10^{-5}$ & 18.9 & 6.6\\
\hline
\end{tabular}
\end{center}
\end{table*}

In Table~\ref{tab:xsect_emu},
the cross section measurements in the \ee\ and \mm\
decay channels are reported in the fiducial volume, which is defined by the lepton
acceptance $\pt^\ell > 20\GeV$ and $|\eta^\ell|<2.4$, and the invariant mass of the lepton
pair $66\GeV<m_{\ell\ell}<116\GeV$.
This implies for the \ee\ decay channel a small acceptance correction due to
the discarded events, in which one or more electrons are within the calorimeter
transition region, $1.37 < |\eta^e| < 1.52$.
The resulting correction of the normalized differential cross section is smaller than $0.6\%$.
The measurement is reported with respect to three distinct reference points at particle level
regarding QED FSR corrections.
The true dilepton mass $m_{\ell\ell}$
and transverse momentum \ptz\ are either defined by the final state leptons
after QED FSR (``bare'' leptons), or by recombining them with radiated photons
within a cone of $\Delta R=0.1$ (``dressed'' leptons), or by the \Zg\
propagator. The propagator definition
corresponds to a full correction for QED FSR effects,
allows for a combination of the electron and muon channels, and facilitates a
direct comparison of the measurement with QCD calculations.
The QED FSR corrections are at most 8\%~(5\%) for the normalized differential
cross section in the \ee~(\mm) decay channel.

\section{Systematic uncertainties}
\label{sec:Systematics}

Systematic uncertainties arise mainly from lepton efficiencies,
momentum scale and resolution, and from the unfolding procedure.
They are evaluated
by varying separately each parameter in question and recalculating the
bin-by-bin correction factors used to unfold the \ptz\ spectrum. The observed deviation
in the unfolded distributions is limited by the statistics of the simulated events.
The statistical component of these deviations is estimated using a
bootstrap method~\cite{Bootstrap} based on multiple resampling of the
simulated events. Each sample
has the same size and may contain the same events
multiple times. Symmetric systematic uncertainties are derived
in such a way~\cite{errorcalc} that the area spanned by these uncertainties covers 68\% of the integral
over a Gaussian with mean and width equal to the mean and standard deviation of
the bootstrap distributions.
The resulting uncertainties are equal to the standard deviation 
if the bootstrap distribution is centred at zero, and approach
the mean plus half the standard deviation if the mean deviates from zero.

The following sources of systematic uncertainties on the measured normalized
differential cross section are evaluated, where the quoted uncertainties are 
relative percentages.

\systHeading{Lepton reconstruction and identification:}
The efficiencies are determined as a function of \ptz\
using a tag-and-probe method. Due to the limited
statistics at high \ptz, the uncertainties are parametrized by a function
which increases with \ptz.
For the \ee\ analysis, the uncertainty due to the electron reconstruction and
identification
varies between 1.0\% and 3.2\%. An additional uncertainty of 0.1\% arises from
the modelling of local calorimeter readout problems.
In the case of the \mm\ selection, the uncertainties due to the 
efficiencies on the trigger, on the muon reconstruction and 
identification, and on the isolation requirement are evaluated 
separately and are found to be within 0.4\%, 2.3\%, and 1.0\%, 
respectively, except for the three highest \ptz\ bins,
where they reach 0.8\%, 4.9\%, and 2.7\%.

\systHeading{Lepton energy (momentum) scale and resolution:} The scale and resolution 
corrections of the simulation are varied within their uncertainties estimated
from the fit to the observed \Zg\ line shape. Correlations across \ptz\ bins 
are taken into account.
Due to the normalization to the inclusive cross section, a
systematic shift at low \ptz\ is balanced by a shift in the opposite
direction at high \ptz.
In the \ee\ (\mm) analysis, the uncertainty due to scale variations
is found to be 2.7\%~(0.2\%) 
for the lowest bin, decreasing down to 0.2\%~($<0.1\%$) at $\ptz\sim 10$~\GeV, and then
increasing up to 4.4\%~(0.4\%) at the highest \ptz\ values.
Uncertainties due to resolution are estimated to be 0.5\% for the \ee\
channel and between 0.1\% and 0.7\% for the \mm\ channel.

\systHeading{Unfolding procedure:}
The bin-by-bin correction factors depend on the shape of the assumed underlying 
\ptz\ distribution, which leads to a systematic uncertainty evaluated in the following way.
The default MC signal sample ({\sc{}Pythia}) is reweighted to different true \ptz\ shapes using {\sc{}Resbos} and \mcatnlo .
The variance of these generator predictions does not entirely cover the 
observed difference between simulation and data. Therefore the {\sc{}Pythia}
signal sample is reweighted, in addition, to distributions based on 
unfolded data.
The spectra obtained by unfolding 
the data either with the bin-by-bin method or alternative matrix unfolding
techniques~\cite{MR99a:65037,Kaipio:1338003} are considered. 
These new corrected spectra feature different uncertainties. 
The bin-by-bin method
suffers a larger systematic uncertainty due to the assumed true \ptz\ shape, 
whereas matrix-based unfolding is nearly independent of this assumption, but 
suffers from a larger statistical uncertainty. 
Each of the reweighted spectra is treated in the same way as the data, 
and is unfolded with
the default bin-by-bin correction factors.
The maximum deviation from the
respective true \ptz\ spectrum is considered as a systematic uncertainty.
A possible influence from the parton shower and hadronization model is 
estimated by comparing the bin-by-bin correction factors determined from 
either the default {\sc{}Pythia} sample or the \mcatnlo\ sample. 
\mcatnlo\ uses {\sc{}Herwig} for the parton shower and {\sc{}Jimmy} for 
the underlying event. 
To separate these model uncertainties from the uncertainty due to the 
underlying \ptz\ distribution, which is already accounted for, the \mcatnlo\ 
sample is reweighted to the \ptz\ shape of the default MC signal sample.
The uncertainties due to the shape of the \ptz\ distribution and due 
to the modelling of the parton shower and the hadronization are combined to 
yield the total unfolding uncertainty,
which is found to be within 2.0\%~(1.3\%) in the \ee\ (\mm) channel
for $\ptz$ between $6\GeV$ and $100\GeV$.
For $\ptz<6\GeV$ the uncertainty is as large as 3.6\%~(4.7\%) and for $\ptz>100\GeV$
it is as large as 4.2\% (2.9\%) in the \ee\ (\mm) channel.
The unfolding uncertainty is dominated by the deviations observed when 
reweighting the default MC signal sample to the \ptz\ distributions 
obtained from the data with the matrix unfolding techniques.

\systHeading{Background contamination:}
Uncertainties in the estimation of the background from QCD multijet, weak
boson, and $t\bar{t}$ production yield values of up to 1.4\%~(0.6\%) for
the \ee\ (\mm) analysis when propagated to the normalized differential
cross section.

\systHeading{Modelling of pileup corrections:}
Pileup has a small influence on this measurement. An uncertainty 
of 0.3\% on the normalized differential cross section is derived.

\systHeading{MC sample statistics:}
The uncertainties are within $0.4\%-1.5\%$~($0.3\%-0.8\%$) in the \ee~(\mm) channel,
except for highest \ptz\ values for which the uncertainties reach 3.6\%~(1.6\%).

\systHeading{QED final state radiation:} A conservative systematic uncertainty of
0.6\% due to the \ptz-dependent modelling of QED FSR is assigned.
This uncertainty addresses both potential differences between the
approximation used in {\sc Photos} compared to exact second order QED FSR matrix
element calculations~\cite{SanC,KKMC} and uncertainties in the simulation of
the interaction of the radiated photons with the detector 
material~\cite{WZXsInclusive}.

The systematic uncertainties listed above are added quadratically to obtain the 
total systematic uncertainties listed in Table~\ref{tab:xsect_emu}.

\section{Results and conclusions}
\label{sec:Results}

\begin{figure}
  \centering
  \includegraphics[width=0.49\textwidth]{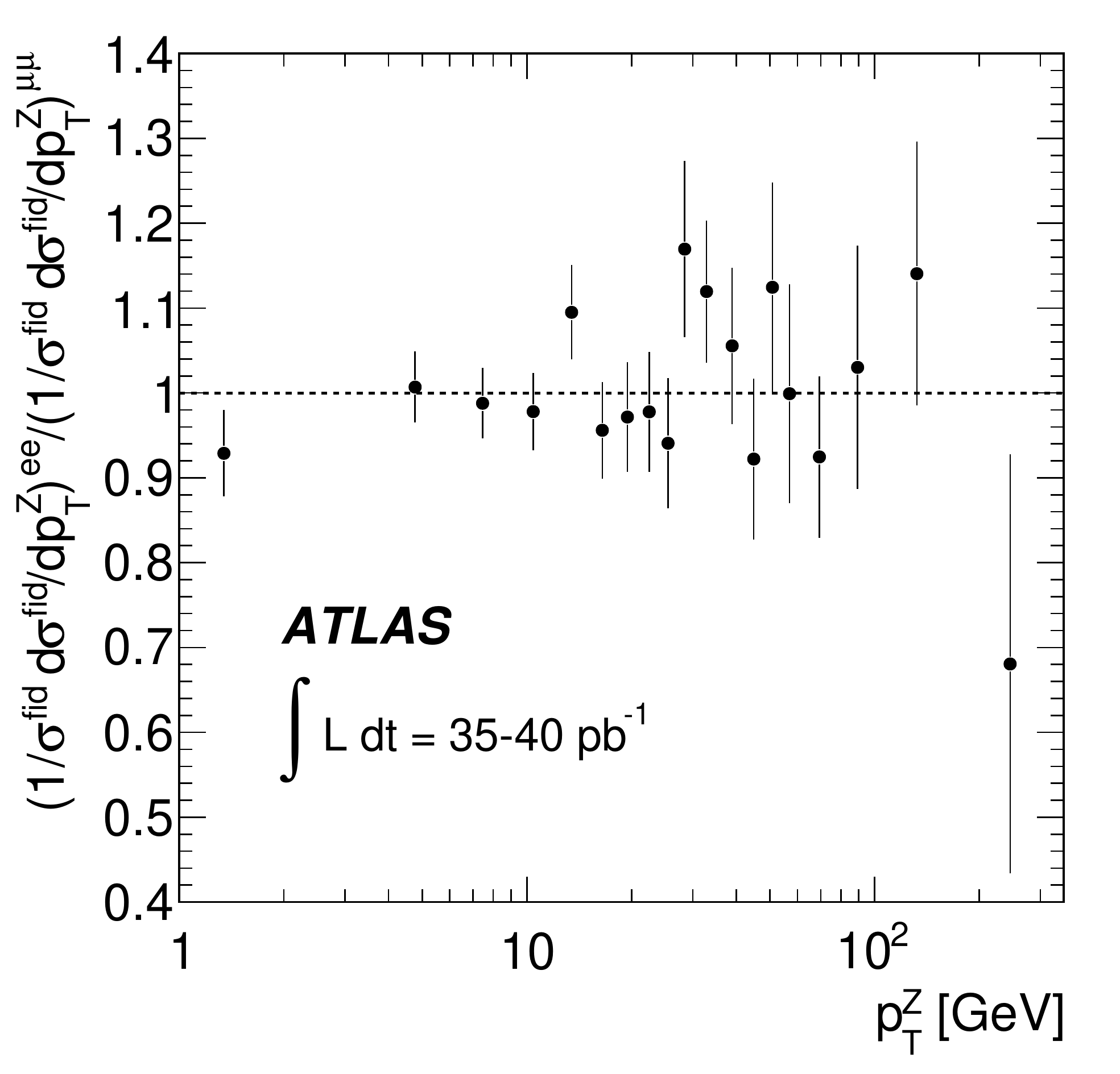}
  \caption{
    Ratio of the normalized differential cross section for 
    \Zgee\ and \Zgmm\ production as a function of
    \ptz\ at \Zg\ propagator level using an integrated luminosity of $35\ \ipb$ and $40\ \ipb$, respectively.
    The error bars shown include statistical and systematic 
    uncertainties. The systematic uncertainties due to the unfolding procedure and QED FSR, which are correlated
    between the electron and muon decay channel, are omitted. 
    \label{fig:ratioemu}
  }
\end{figure}

The normalized differential cross section
measurements for \Zgee\ and \Zgmm\ production are in good agreement with each other at
the \Zg\ propagator level; see 
Figure~\ref{fig:ratioemu} for the ratio of the measured cross sections.
The two decay channels are combined at \Zg\ propagator level
using a $\chi^2$ minimization method which takes
into account the correlated systematic uncertainties for the
\ee\ and \mm\ channels~\cite{H1:2009bp}. 
The uncertainties due to the unfolding procedure and QED FSR are
considered to be common for the two channels. The minimization 
yields a $\chi^2/\dof=17.0/19$ indicating the excellent 
compatibility of the electron and muon data.

\begin{figure*}
  \centering
  \begin{overpic}[width=0.49\textwidth]{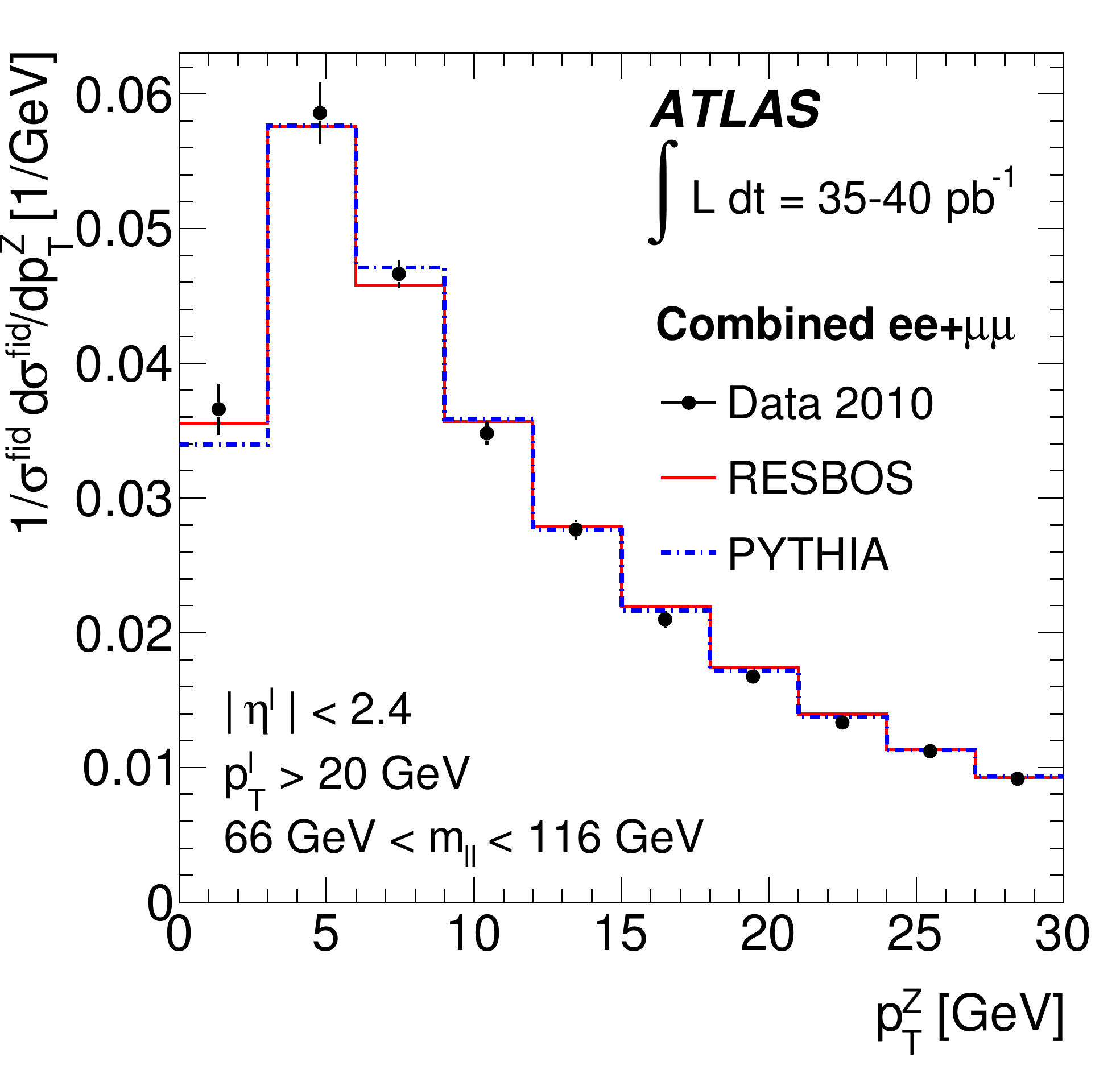}
    \put(84,84.5){\large(a)}
  \end{overpic}
  \begin{overpic}[width=0.49\textwidth]{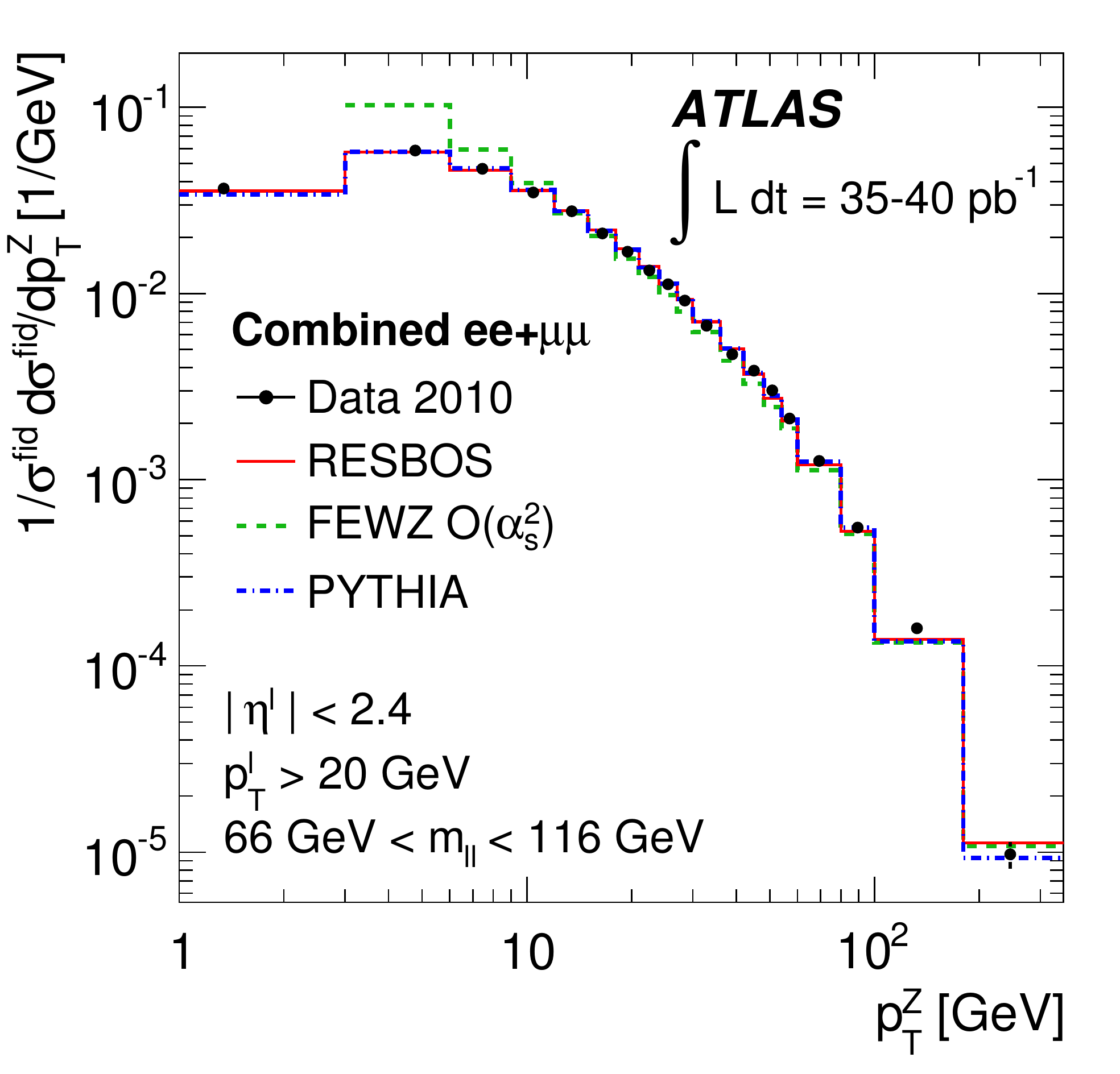}
    \put(84,84.5){\large(b)}
  \end{overpic}

  \caption{
    The combined normalized differential cross section at \Zg\ propagator 
    level as a function of
    \ptz\ for (a) the range $\ptz <30$~\GeV\ and (b) the full range
    compared to the predictions of {\sc Resbos}, {\sc Pythia}, and 
    {\sc Fewz} at $\mathcal{O}(\alphas^2)$. The error bars shown include statistical 
    and systematic uncertainties. For the combination, the \ee\ (\mm) channel contributes
    with an integrated luminosity of $35\ \ipb$ ($40\ \ipb$). At low \ptz\ the {\sc Fewz} 
    prediction diverges and is omitted.
    \label{fig:sigma}
  }
\end{figure*}

The combined measurement of the normalized differential cross section within
the fiducial lepton acceptance as a function of \ptz,  
$1/\sigma^{\mathrm{fid}} \der\sigma^{\mathrm{fid}}/\der \ptz$,
is shown in Figure~\ref{fig:sigma} and Table~\ref{tab:xsect_comb}.
In addition, the acceptance corrections $A_c$ needed to extrapolate the
measurement to full lepton acceptance, but keeping the mass range
$66\GeV<m_{\ell\ell}<116\GeV$, are reported. They are determined
using {\sc Pythia} and the MRST2007LO$^*$ PDF set. 
The acceptance for the inclusive fiducial cross section is $0.48$. However, the
acceptance corrections $A_c$ for the normalized differential cross
section are within $10\%$ of $1.0$ for the bins with $\ptz<80\GeV$.
The uncertainty on $A_c$ is estimated by: reweighting the {\sc Pythia}
prediction using the HERAPDF1.0~\cite{Aaron:2009wt} and 
CTEQ6.6~\cite{Nadolsky:2008zw} parton distribution functions;
propagating the CTEQ6.6 PDF error eigenvector sets; and by taking into 
account the difference to the
predictions obtained with \mcatnlo, {\sc Resbos}, and {\sc Fewz}.

\begin{table}[tb]
\caption{
The combined normalized differential cross section at \Zg\ propagator level,
$1/\sigma^{\mathrm{fid}}\der\sigma^{\mathrm{fid}}/\der \ptz$, 
as a function of the average \Zg\ transverse momentum $\langle\ptz\rangle$
with relative statistical and total systematic uncertainties. The 
multiplication with the inverse acceptance correction $A^{\rm -1}_c$ (given with
uncertainties, ``unc.'') yields the normalized differential cross section 
$1/\sigma^{\mathrm{tot}}\der\sigma^{\mathrm{tot}}/\der \ptz$ extrapolated to the full lepton
acceptance. The data can be obtained electronically through the HepData repository~\cite{hepdata}.
\label{tab:xsect_comb}
}
\begin{center}
\begin{tabular}{crrrrr}
\hline\noalign{\smallskip}
\multicolumn{1}{c}{$\langle\ptz\rangle$} & 
\multicolumn{1}{c}{$\frac{1}{\sigma^{\mathrm{fid}}} \frac{\der\sigma^{\mathrm{fid}}}{\der \ptz}$} &
\multicolumn{1}{c}{stat.} & \multicolumn{1}{c}{syst.} & \multicolumn{1}{c}{$A^{\rm{-1}}_c$} & \multicolumn{1}{c}{unc.} \\
\multicolumn{1}{c}{(\GeV)} & \multicolumn{1}{c}{(\GeV$^{-1}$)} & \multicolumn{1}{c}{(\%)} & \multicolumn{1}{c}{(\%)} & & \multicolumn{1}{c}{(\%)}\\
\hline\noalign{\smallskip}
1.3 & 0.0366 & 2.0 & 4.7 & 1.047 & 3.7\\
4.8 & 0.0586 & 1.5 & 3.6 & 1.029 & 1.8\\
7.5 & 0.0466 & 1.7 & 1.5 & 1.014 & 1.5\\
10  & 0.0348 & 1.9 & 1.6 & 0.999 & 1.5\\
13  & 0.0277 & 2.2 & 1.7 & 0.999 & 1.4\\
16  & 0.0210 & 2.5 & 1.7 & 0.990 & 1.5\\
19  & 0.0167 & 2.8 & 1.8 & 0.989 & 1.5\\
22  & 0.0133 & 3.1 & 1.9 & 0.990 & 1.5\\
25  & 0.0112 & 3.4 & 2.0 & 0.994 & 2.3\\
28  & 0.0092 & 4.0 & 2.1 & 0.988 & 2.3\\
33  & 0.0067 & 3.2 & 2.1 & 0.987 & 3.2\\
39  & 0.0047 & 3.8 & 2.3 & 0.979 & 3.9\\
45  & 0.0038 & 4.2 & 2.4 & 0.965 & 4.3\\
51  & 0.0030 & 4.9 & 2.5 & 0.950 & 4.4\\
57  & 0.0021 & 5.7 & 2.7 & 0.938 & 5.3\\
69  & 0.0013 & 4.0 & 2.8 & 0.910 & 5.3\\
89  & $5.5\cdot 10^{-4}$ & 6.1 & 3.1 & 0.894 & 5.3\\
132 & $1.6\cdot 10^{-4}$ & 5.9 & 3.7 & 0.826 & 5.4\\
245 & $9.8\cdot 10^{-6}$ & 15.6 & 5.4 & 0.672 & 5.6\\
\hline
\end{tabular}
\end{center}
\end{table}

\begin{figure*}
  \centering
  \begin{overpic}[width=0.49\textwidth]{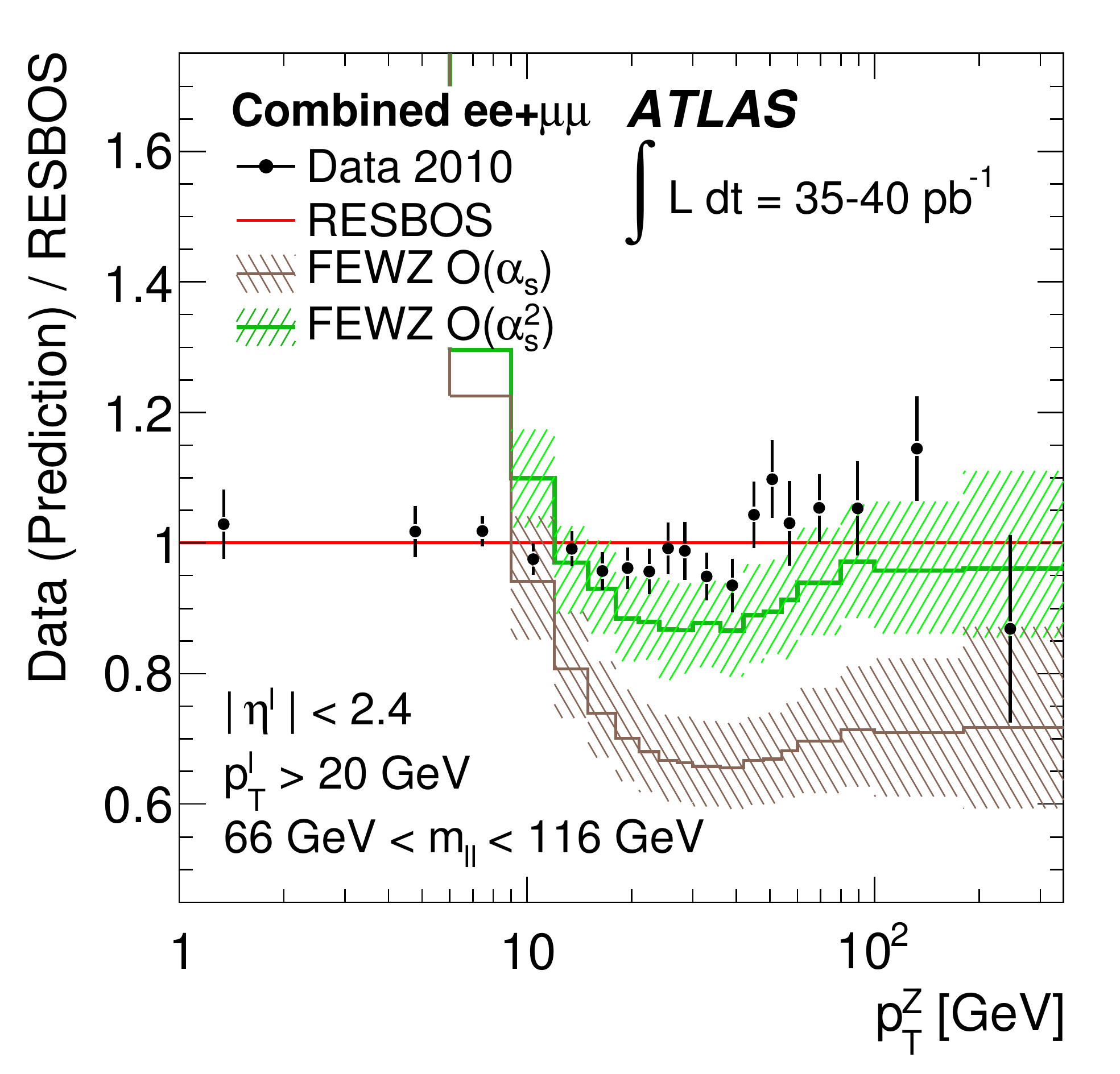}
    \put(84,84.5){\large(a)}
  \end{overpic}
  \begin{overpic}[width=0.49\textwidth]{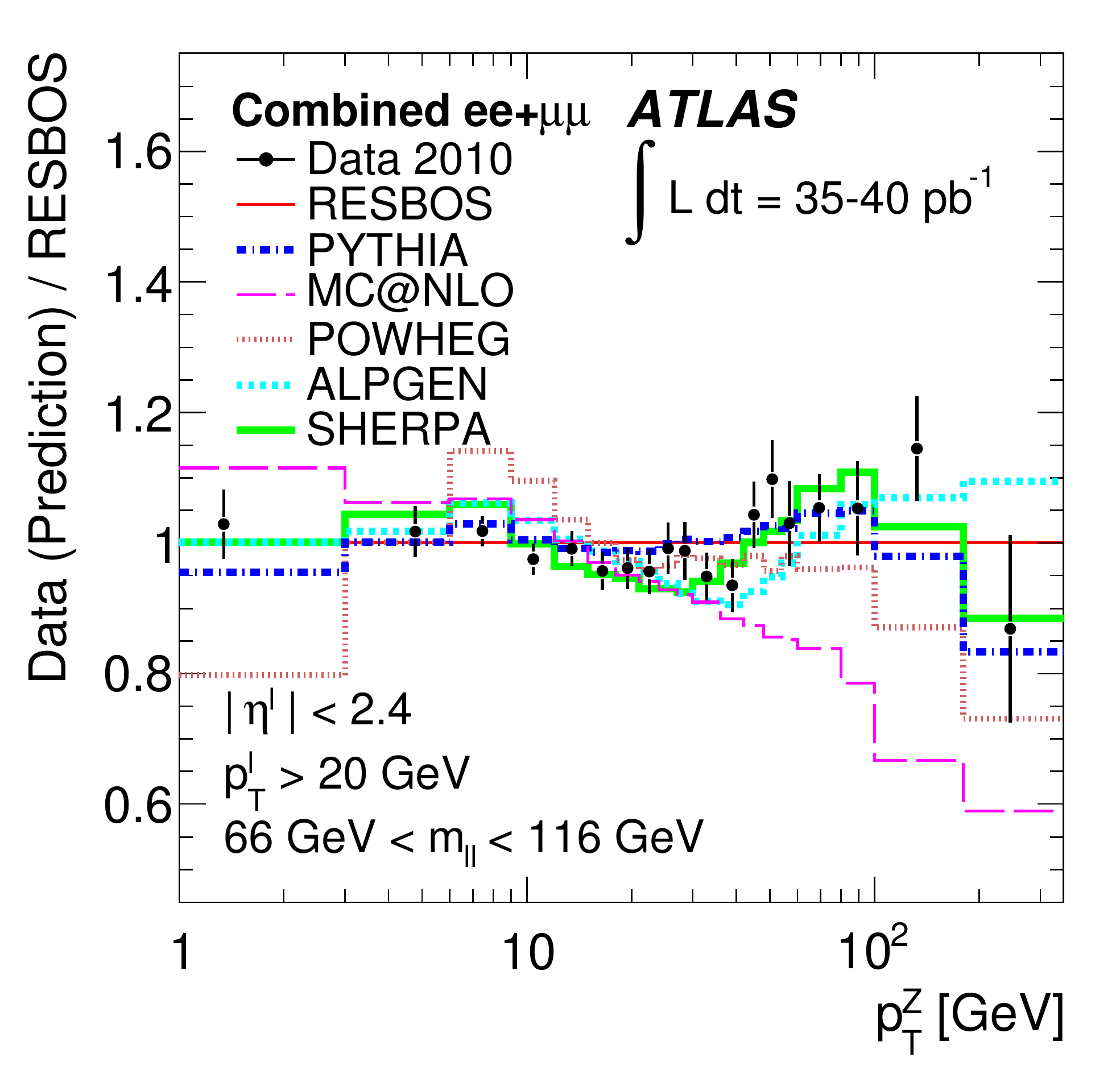}
    \put(84,84.5){\large(b)}
  \end{overpic}

  \caption{
    Ratios of the combined data and various predictions over the {\sc{}Resbos} prediction
    for the normalized differential 
    cross section as a function of \ptz: (a) {\sc Fewz} predictions
    at $\mathcal{O}(\alphas)$ and $\mathcal{O}(\alphas^2)$;
    (b) predictions from the generators {\sc Pythia}, \mcatnlo, {\sc Powheg}, 
    {\sc Alpgen} and {\sc Sherpa}.
    The {\sc Fewz} predictions are shown with combined scale, \alphas, and
    PDF uncertainties. The data points are shown with combined statistical and systematic
    uncertainty. At low \ptz\ the $\mathcal{O}(\alphas)$ and $\mathcal{O}(\alphas^2)$ predictions of {\sc{}Fewz} diverge and are
    omitted. 
    \label{fig:ratio}
  }
\end{figure*}

In Figure~\ref{fig:ratio}, the measurement within the fiducial 
acceptance is compared with predictions of pQCD calculations and of 
various event generators introduced above.
The $\mathcal{O}(\alphas)$ and $\mathcal{O}(\alphas^2)$ pQCD predictions
of the \ptz\ dependent cross section are obtained with {\sc Fewz} v2.0~\cite{Gavin:2010az} 
and the MSTW2008
PDF sets~\cite{Martin:2009iq}. The inclusive cross section, which is used
to normalize the prediction, is calculated in the same way.
The uncertainties on the normalized predictions are evaluated by
variation of the renormalization and factorization scales by factors of two 
around the nominal scale $\mu_R=\mu_F=M_Z$ with the constraint
$0.5\le \mu_R/\mu_F \le 2$, by variation of \alphas\ within a range
corresponding to 90\% confidence-level limits~\cite{Martin:2009bu},
and by using the PDF error eigenvector sets at 90\% confidence-level. They amount to
$\sim10\%$ and $\sim8\%$ for the $\mathcal{O}(\alphas)$ and 
$\mathcal{O}(\alphas^2)$ prediction, respectively, with a dominant contribution 
of 9\% and 6.5\% from the scale variations. 
In contrast to the $\Zg$ inclusive cross section, the prediction of the \ptz\
distribution suffers from substantial scale uncertainties indicating 
non-negligible missing higher order corrections.
For $\ptz>18$~\GeV, the pQCD prediction receives an $\mathcal{O}(\alphas^2)$ correction of
$26-36\%$. Despite this correction, 
the $\mathcal{O}(\alphas^2)$ prediction undershoots the data by about 10\%, which is comparable to the size of
the scale uncertainty. This deficit is smaller compared to the $15-25\%$
difference observed at the Tevatron~\cite{Abazov:2007nt,Abazov:2010kn}.
At low boson transverse momenta, where fixed order pQCD calculations
are not expected to give an adequate description of the cross section,
the disagreement increases rapidly towards vanishing \ptz.

In addition, the measurement is compared to the predictions of {\sc{}Resbos} and
various event generators. The consistency with the data is verified with a
$\chi^2$ test which uses the $\chi^2$ definition also used for the combination of
the \ee\ and \mm\ decay channels.


The {\sc Resbos}~\cite{Balazs:1997xd} prediction, which combines
resummed and fixed order pQCD calculations, 
is based on the CTEQ6.6~\cite{Nadolsky:2008zw} PDF set and a resummation 
scale of $m_{Z}$. 
It is verified that the different PDF sets
used for the {\sc Fewz} and {\sc{}Resbos} predictions 
lead to differences below 3\%.
{\sc Resbos} shows good agreement with 
the measurement over the entire \ptz\ range ($\chi^2/\dof=21.7/19$), 
indicating the importance of resummation even at relatively large \ptz.
However, its predictions are slightly higher than the data for \ptz\ values in the range of $10\GeV$ to $40\GeV$ and
slightly lower above $40\GeV$. 

The {\sc Alpgen}~\cite{Mangano:2002ea} and {\sc Sherpa}~\cite{Gleisberg:2008ta} generators consider processes with up to five 
additional hard partons associated with the produced boson and give a
good description of the entire measured spectrum, up to large \ptz, with $\chi^2/\dof$
of $31.9/19$ and $16.8/19$, respectively. 
Here, the enhancement of the cross section compared to the 
$\mathcal{O}(\alphas^2)$ prediction can be attributed to processes with
large parton multiplicities~\cite{Krauss:2005nu}, which correspond to tree-level diagrams of higher
order in the strong coupling. 
{\sc Sherpa} v1.2.3 and {\sc Alpgen} v2.13 are used, with the latter
being interfaced 
to {\sc Herwig} v6.510~\cite{herwig} for parton shower and fragmentation into 
particles, and to {\sc Jimmy} v4.31~\cite{Butterworth:1996zw} to model 
underlying event contributions.
For {\sc Alpgen}, the CTEQ6L1~\cite{Pumplin:2002vw} PDF set is employed 
and the factorization scale is set to $\mu_F^2=m_{\ell\ell}^2+\sum \pt^2$, 
where the sum extends over all associated partons.
The {\sc Sherpa} prediction uses the CTEQ6.6 PDF set and $\mu_F^2=m_{\ell\ell}^2+{(\ptz)}^2$.

The predictions of the parton shower event generators {\sc Pythia}
and \mcatnlo\ are based on the simulated samples as described above. Figure~\ref{fig:ratio}b also shows
the predictions of {\sc{}Powheg} v1.0~\cite{powheg} interfaced to a {\sc{}Pythia} version with an underlying event tune  to
Tevatron and $7\TeV$ $pp$ collision data~\cite{mc10Pythia}. Whereas
\mcatnlo\ ($\chi^2/\dof = 111.6/19$) and {\sc{}Powheg} ($\chi^2/\dof = 100.4/19$) 
deviate from the data at low and high \ptz, {\sc Pythia}
describes the measurement well over the entire range of boson transverse 
momentum ($\chi^2/\dof=17.9/19$). 

In summary, the \Zg\ transverse momentum differential distribution
has been measured up to $\ptz=350$~\GeV\
for electron and muon pairs with invariant masses $66\GeV<m_{\ell\ell}<116\GeV$
produced in $pp$ collisions at $\sqrt{s}=7$~TeV
based on integrated luminosities of 35~\ipb\ and 40~\ipb, respectively,
recorded with the ATLAS detector.
{\sc Resbos} describes the spectrum well for
the entire \ptz\ range. At $\ptz>18\GeV$, the central {\sc{}Fewz} $\mathcal{O}(\alphas^2)$ prediction
underestimates the data by about 10\%, which is comparable to the size of
the combined experimental and theoretical uncertainty. 
The measurement is compared to 
predictions of various event generators and a good agreement with
{\sc{}Sherpa}, {\sc{}Alpgen}, and {\sc{}Pythia} is found. 
Except for the lowest \ptz\ values, the measurement
is limited by statistics rather than systematic uncertainties.
The systematic uncertainties are also mostly limited by the size of the data 
sample and are expected to improve with increasing integrated luminosity.

\global\emergencystretch = .2\hsize

\section{Acknowledgements}

We thank CERN for the very successful operation of the LHC, as well as the
support staff from our institutions without whom ATLAS could not be
operated efficiently.

We acknowledge the support of ANPCyT, Argentina; YerPhI, Armenia; ARC,
Australia; BMWF, Austria; ANAS, Azerbaijan; SSTC, Belarus; CNPq and FAPESP,
Brazil; NSERC, NRC and CFI, Canada; CERN; CONICYT, Chile; CAS, MOST and
NSFC, China; COLCIENCIAS, Colombia; MSMT CR, MPO CR and VSC CR, Czech
Republic; DNRF, DNSRC and Lundbeck Foundation, Denmark; ARTEMIS, European
Union; IN2P3-CNRS, CEA-DSM/IRFU, France; GNAS, Georgia; BMBF, DFG, HGF, MPG
and AvH Foundation, Germany; GSRT, Greece; ISF, MINERVA, GIF, DIP and
Benoziyo Center, Israel; INFN, Italy; MEXT and JSPS, Japan; CNRST, Morocco;
FOM and NWO, Netherlands; RCN, Norway; MNiSW, Poland; GRICES and FCT,
Portugal; MERYS (MECTS), Romania; MES of Russia and ROSATOM, Russian
Federation; JINR; MSTD, Serbia; MSSR, Slovakia; ARRS and MVZT, Slovenia;
DST/NRF, South Africa; MICINN, Spain; SRC and Wallenberg Foundation,
Sweden; SER, SNSF and Cantons of Bern and Geneva, Switzerland; NSC, Taiwan;
TAEK, Turkey; STFC, the Royal Society and Leverhulme Trust, United Kingdom;
DOE and NSF, United States of America.

The crucial computing support from all WLCG partners is acknowledged
gratefully, in particular from CERN and the ATLAS Tier-1 facilities at
TRIUMF (Canada), NDGF (Denmark, Norway, Sweden), CC-IN2P3 (France),
KIT/GridKA (Germany), INFN-CNAF (Italy), NL-T1 (Netherlands), PIC (Spain),
ASGC (Taiwan), RAL (UK) and BNL (USA) and from the Tier-2 facilities
worldwide.

\bibliographystyle{model1-num-names}
\bibliography{ZpTbibliography}

\onecolumn
\begin{flushleft}
{\Large The ATLAS Collaboration}

\bigskip

G.~Aad$^{\rm 48}$,
B.~Abbott$^{\rm 111}$,
J.~Abdallah$^{\rm 11}$,
A.A.~Abdelalim$^{\rm 49}$,
A.~Abdesselam$^{\rm 118}$,
O.~Abdinov$^{\rm 10}$,
B.~Abi$^{\rm 112}$,
M.~Abolins$^{\rm 88}$,
H.~Abramowicz$^{\rm 153}$,
H.~Abreu$^{\rm 115}$
% ...
\bigskip

% ...
$^{10}$ Institute of Physics, Azerbaijan Academy of Sciences, Baku, Azerbaijan\\
$^{11}$ Institut de F\'isica d'Altes Energies and Universitat Aut\`onoma  de Barcelona and ICREA, Barcelona, Spain\\
% ...
$^{48}$ Fakult\"{a}t f\"{u}r Mathematik und Physik, Albert-Ludwigs-Universit\"{a}t, Freiburg i.Br., Germany\\
$^{49}$ Section de Physique, Universit\'e de Gen\`eve, Geneva, Switzerland\\
% ...
$^{88}$ Department of Physics and Astronomy, Michigan State University, East Lansing MI, United States of America\\
% ...
$^{111}$ Homer L. Dodge Department of Physics and Astronomy, University of Oklahoma, Norman OK, United States of America\\
$^{112}$ Department of Physics, Oklahoma State University, Stillwater OK, United States of America\\
% ...
$^{115}$ LAL, Univ. Paris-Sud and CNRS/IN2P3, Orsay, France\\
% ...
$^{118}$ Department of Physics, Oxford University, Oxford, United Kingdom\\
% ...
$^{153}$ Raymond and Beverly Sackler School of Physics and Astronomy, Tel Aviv University, Tel Aviv, Israel\\
% ...
\end{flushleft}


\end{document}